# Polymer-embedded molecular junctions between graphene sheets: A Molecular Dynamics study.

Alessandro di Pierro and Alberto Fina

Dipartimento di Scienza Applicata e Tecnologia, Politecnico di Torino, Alessandria Campus,

Viale Teresa Michel 5, 15121 Alessandria, Italy

Corresponding author: alberto.fina@polito.it

## Abstract
Thermal conduction in polymer nanocomposites depends on several parameters including the thermal conductivity and geometrical features of the nanoparticles, the particle loading, their degree of dispersion and formation of a percolating networks. To enhance efficiency of thermal contact between free-standing conductive nanoparticles were previously proposed. This work report for the first time the investigation of molecular junctions within a graphene polymer nanocomposite. Molecular dynamics simulations were conducted to investigate the thermal transport efficiency of molecular junctions in polymer tight contact, to quantify the contribution of molecular junctions when graphene and the molecular junctions are surrounded by polydimethylsiloxane (PDMS). A strong dependence of the thermal conductance in PDMS/graphene model was found, with best performances obtained with short and conformationally rigid molecular junctions.

## Methods
**Non-Equilibrium Molecular Dynamics (NEMD)**

To evaluate the thermal properties of condensed matter, Non-Equilibrium Molecular Dynamics (NEMD) approach is an established technique used to investigate the thermal property of materials [1]. In NEMD, the basic idea is to create a thermal gradient inside the material and measure the derived heat flux. The temperature gradient arises by the application of thermostats to limited regions of the domain (Figure 1). The regions excluded from the thermostats run inside NVE statistical ensemble, experiencing the temperature difference, according to the model design. All the simulations that involved only alkyl chains between the graphene sheets adopted the AIREBO force field, while for the systems which required various chemical species, the COMPASS force field was employed.

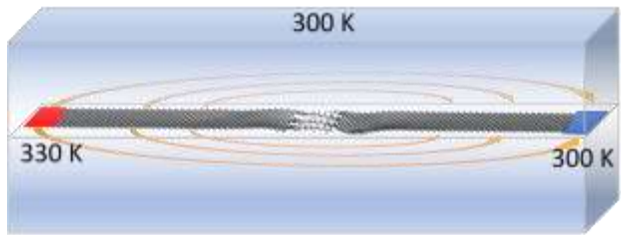

*Figure 1. The typical layout adopted in NEMD simulations.*

The simulation procedure followed a stepwise scheme: at the beginning, the whole system was initially relaxed through an equilibration period of at least 125 ps in canonical ensemble at 300 K, a temperature in which the quantum effect is considered negligible [2]. Then, the ends of the model were fixed in an 8 Å long region and the rest of the model is virtually split into 22 thermal layers along the x coordinate to avoid sliding and PBC interference. In this stage, two thermostats are applied to the first and last thermal layers [3] between the fixed atoms region. At the end of the thermal equilibration, followed a minimum 250 ps of thermo-stated preheating. In this stage, Nosé-Hoover thermostats were set to 310 K and 290 K and applied to the model ends. All the atoms in the system, except those in thermo-stated regions, run in microcanonical ensemble and a thermal gradient is gradually established inside the model. With the system in steady state, data collection [4, 5] of energy and temperature was performed for at least 4 ns. The thermal flow through thermostats was is calculated from the slope the energy versus time plot [3], while the group temperatures were computed from the averaging the instantaneous local kinetic temperature [4, 5] as reported in Equation 1.

$$T_i = \frac{2}{3 N_i k_B} \sum_j \frac{p_j^2}{2 m_j} \qquad 1$$

In Equation 1, $T_i$ is the temperature of $i^{th}$ group of atoms, $N_i$ is the number of atoms in $I^{th}$ group, $k_B$ is the Boltzmann's constant, $m_j$ and $p_j$ are atomic mass and momentum of atom $j$, respectively. The temperature of each slab is computed by a time averaging operation along the simulation time.

Within this method, longer simulation time assures higher accuracy of the temperature calculation. From the plot of the averaged temperature of the slabs, as a function of the coordinate displacement, it is possible to evaluate the thermal jump across the junction. The thermal conductance is then computed through Equation 2,

$$G_c = \frac{\dot{q}}{A \cdot \Delta T} \qquad 2$$

where $G_c$ is the thermal boundary conductivity, $\dot{q}$ the heat flux flowing into the material, A the interface area and ΔT the temperature across the interface.

**Polymer modelling**

PolyDiMethyl Siloxane (PDMS) modelling started by the design of a 49 monomer single chain, Si-methyl terminated, for a sum of 507 atoms (Figure 2). All bonds were described using COMPASS force field.

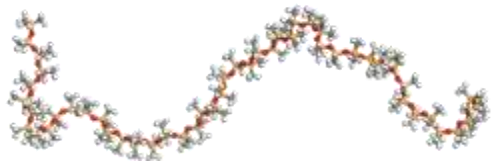

*Figure 2. Representation of a PDMS single chain. Color codes: carbon in grey, hydrogen in white, oxygen in red and silicon in yellow.*

NPT up to 15 K atmospheres for 500 ps. Such pressure was then maintained constant for 500 ps. After this stage, the pressure was reduced to about 5 K atmospheres to reach final density of the polymer of 0.97 (±0.05) g cm$^{-3}$

The final density of the polymer volume was set to 0.97 (±0.05) g cm$^{-3}$, to fit the typical values reported in the previous literature[6].

**Molecular junctions in polymer matrix**

Three different linkers, from the previous studied ones, were selected and incorporated in PDMS. As a first case study, an unjointed model made of two graphene sheets was designed to represent a classical "contact" between graphene flakes inside a polymer composite, referred as "No Linkers" and depicted in Figure 3A. The first molecular junction is the aliphatic/aromatic C5OP, Figure 3B. The second molecule is biphenyl (BP, Figure 3C), which represents a short aromatic junction, and third, the anthracene (ACN, Figure 3D), as theoretical upper bound junction.

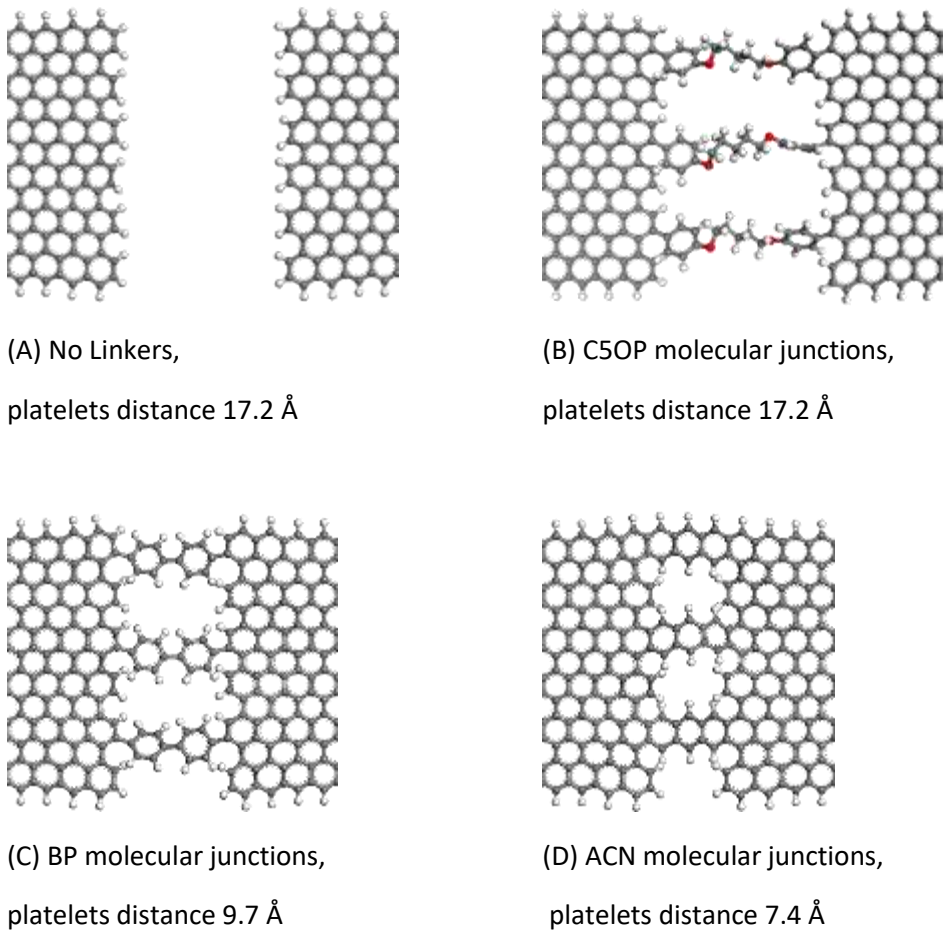

(A) No Linkers,

platelets distance 17.2 Å

(B) C5OP molecular junctions,

platelets distance 17.2 Å

(C) BP molecular junctions,

platelets distance 9.7 Å

(D) ACN molecular junctions,

platelets distance 7.4 Å

*Figure 3. Junction details in polymer bound models. Color codes: carbon atoms in grey, oxygen in red, hydrogen in white.*

Independently on the presence and type of molecular junctions, the system layout is based on a graphene sandwich between two blocks made of PDMS molecules (Figure 4).

The specifications of the model with embedded junction were:

- Two 95 x 25 Å² graphene platelets in armchair configuration
- Two 220 x 35 Å² polymer blocks above and below the graphene platelets and junction region.
- Three molecular junctions (for C5OP or BP or ACN models)

Overall, about 37000 atoms constitutes the assembly of polymer, graphene and thermal linkers. Figure 4 depicts the typical initial model layout, from which flakes, junctions and polymer are clearly distinguishable.

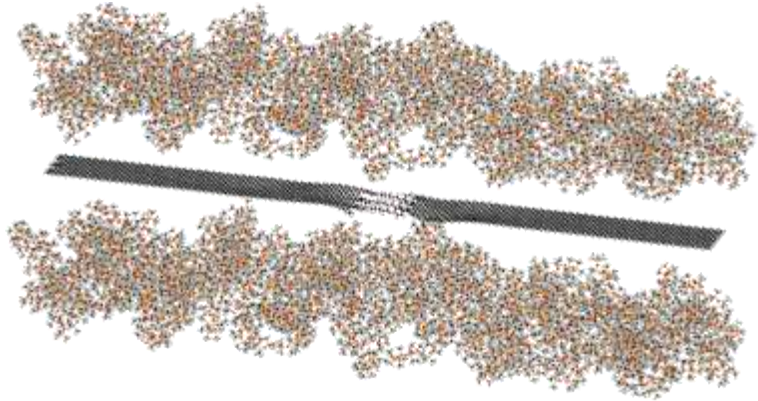

*Figure 4. Initial design view of graphene flakes joint by three C5OP linkers with blocks of PDMS molecules above and below the junction. Color codes: carbon atoms in black, oxygen in red, hydrogen in light gray and silicon in yellow.*

The thermal transport across graphene platelets and molecular junctions was determined through NEMD calculations. At the beginning of the simulation, a 500 ps NPT high-pressure stage was performed to relax the structure and bring the system close to the PDMS actual density, allowing the PDMS polymer and the graphene junction to became a dense system, where the graphene junction is embedded in the surrounding polymer. The picture of Figure 5 represent how the polymer embeds the graphene junction. It is worth noting that the graphene sheets are not flat after the compression stage. This is due to the strong interaction of graphene with PDMS [7], where the larger mass of PDMS forces the thin graphene sheets (including junction) to fold and crease. This is fact realistic and well representative of actual flexible graphene sheets dispersed in a polymer matrix [8], as typically observed through transmission electron microscopy [9-11].

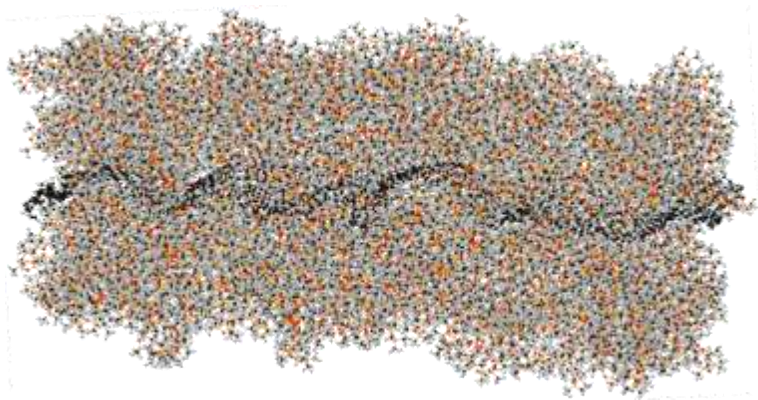

*Figure 5. PDMS polymer surrounding graphene flakes jointed by three C5OP linkers (not clearly visible owing to the presence of polymer). Color codes: Carbon atoms in black, Oxygen in red, Hydrogen in light gray and silicon in yellow. The simulation box size is about 180x75 Å².*

The application of NEMD method began with 1 ns of non-equilibrium pre-heating, followed 2.5 ns of simulation time, where energy and temperature information were collected. For this layout, two thermal layers, corresponding to the graphene sheets ends, acted as thermostats at 300 K and 330 K. The rest of the model, including the PDMS atoms, run in microcanonical ensemble with 300 K of initial temperature.

In suspended graphene junctions, the thermal linkers were the only possible paths from which thermal transport could occur between the hot and the cold ends of the system. The presence of polymer above and below the graphene platelets allows additional heat transfer modes, driven by VdW interaction. These include:

- Heat transfer between the graphene surface and the PDMS parallel to graphene flakes. The large contact area made by graphene and polymer, from the upper and the lower surface of both flakes (corresponding to about 9500 Å$^2$) drains significant heat flux from the hot graphene flake to the matrix and from the matrix to the cold graphene.

- Transfer between molecular junctions and the surrounding polymer molecules. This mode is due to presence of molecular junctions.

- Transfer by molecules trapped between the graphene edges in the center of the model. In this case, PDMS molecules acts in series with graphene edges. This mode is relevant considering the creased shape that graphene assumes during simulation.

## Results and Discussion

For the four different PDMS-embedded molecular junctions described above, the energy transferred from the hot thermostat region to the cold thermostat region is reported in Figure 6. The plots for the upper branch, corresponding to the injected heat from heat source, and the lower branch, which represents the drained energy from the heat sink, appear approx. linear and with limited slope differences, within ± 10% from the average value, as reported in Table 1. The energy flux slopes herein are obtained for the entire set of replicas, each adopting a different velocities seed. Within the same table, the uncertainty from differences in slope between the upper and lower branches (thermostats error) and the uncertainty from replication (seed error) are reported. Moreover, the last column of Table 1 indicates the ratio between the heat flux enhancements from the linker adoption, over the linker-less condition.

The amount of the transferred heat reported in Table 1 indicates that the presence of molecular linkers induces an overall improvement in thermal transport between the graphene flakes. In fact, the aliphatic/aromatic C5OP molecular junction determines a moderate increase in heat flux, rising from 1.47 to 1.53 eV ps$^{-1}$ with an average improvement of about 4 %. For the aromatic biphenyl, the heat flux rises up to 1.65 eV ps$^{-1}$ and for the anthracene, a value of 1.78 eV ps$^{-1}$ was calculated, with a gain of about 12 % and 21 % respectively.

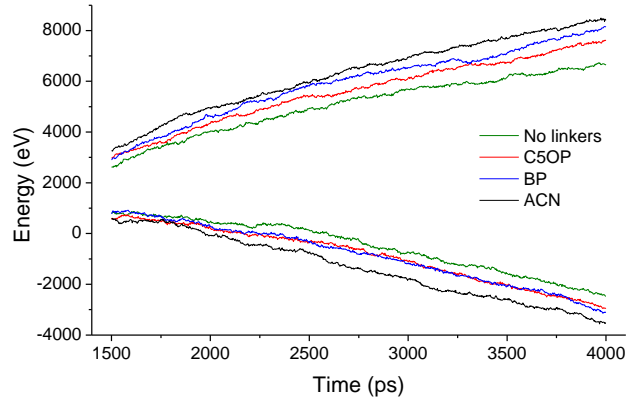

*Figure 6. Energy flowing through thermostats as a function of the time for PDMS-surrounded graphene interfaces. Four simulations are taken as examples for three different linkers (C5OP, BP, and ACN) and a linker-less model (No linkers).*

*Table 1. Averaged Heat flux for different linkers in PDMS enclosed models. Average value, thermostat error, seed error from replicas and linker gain as the ratio of linker heat flux over linker-less heat flux value.*

| Model | Energy flux [eV/ps] | | | Linker gain [%] |
|---|---|---|---|---|
| | Average value | Thermostat error $\frac{\max - \min}{2}$ | Seed error $\frac{\max - \min}{2}$ | $\frac{\text{linkers} - \text{no linkers}}{\text{no linkers}}$ |
| No linkers | 1.47 | ± 0.07 | ± 0.02 | - |
| C5OP | 1.53 | ± 0.12 | ± 0.09 | 4.0 |
| BP | 1.65 | ± 0.14 | ± 0.08 | 12.2 |
| ACN | 1.78 | ± 0.11 | ± 0.06 | 21.1 |

The temperatures of the graphene thermal layers for PDMS-embedded junctions investigated are reported in Figure 7, where empty dots indicated the points excluded from the fitting operation, as the one close to thermostats or across the junction.

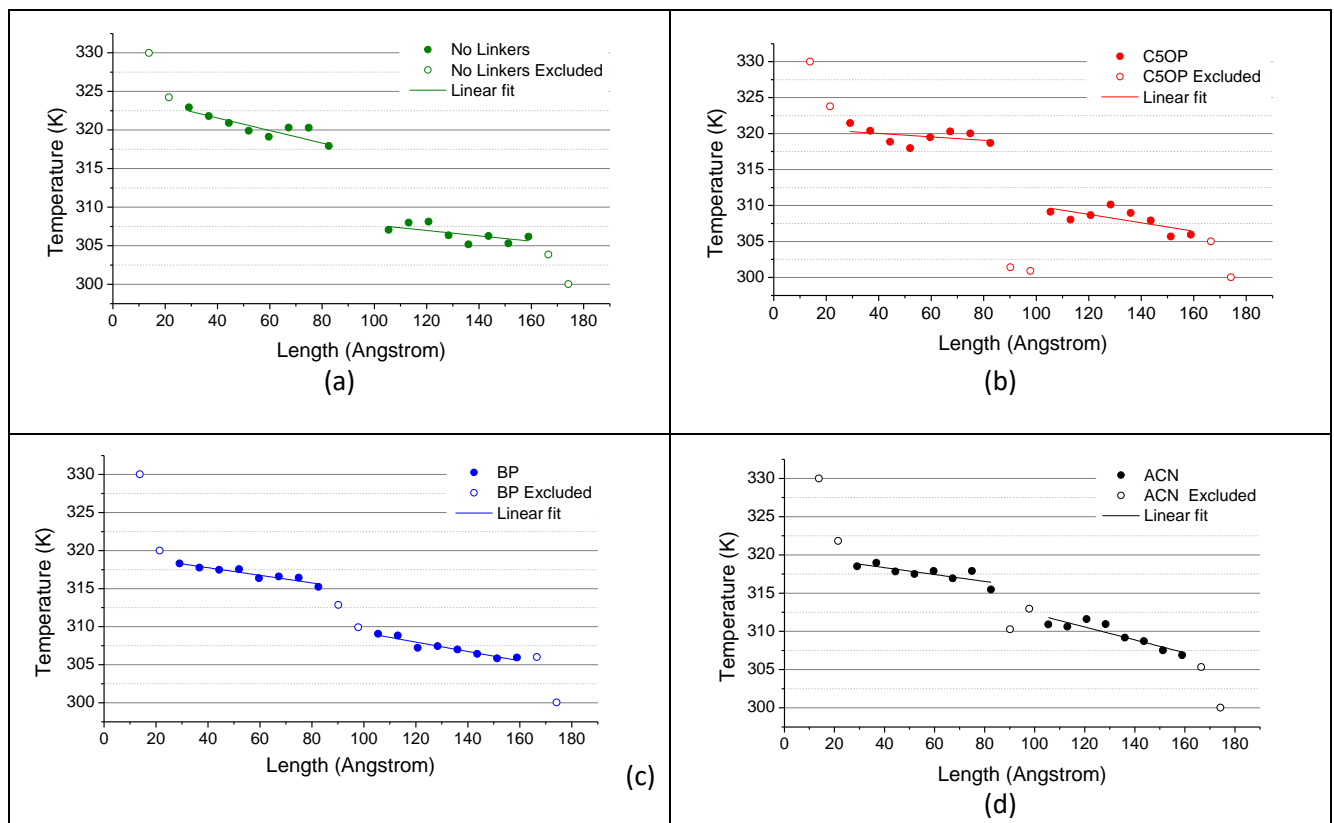

*Figure 7. Temperature as a function of the position of thermal slabs in the absence of molecular junction (a), with C5OP (b), with biphenyl (c) anthracene (d) junctions, embedded in PDMS.*

The model with no molecular junction between graphene foils (Figure 7a), reported an average temperature jump, evaluated from the projection of the slopes of about 8.65 ± 0.57 K. By the heat flux calculation reported in Table 1 and the temperature difference across the junction (Table 2), the thermal conductance reported a value of about 288 ± 21 MW m$^{-2}$ K$^{-1}$. Is noteworthy that all the conductance values reported herein are calculated, considering the whole interface between graphene flakes and PDMS as contact area (9500 Å$^2$), thus representing the conductance of the whole system between the two thermostats as a contribution of ITC between graphene and PDMS and TBC trough edges. Adopting the same methodology for the other models where linkers were employed, the C5OP temperature jump (Figure 7b and Table 2) was calculated of about 7.46 ± 0.31 K and consequently the thermal conductance increased to 346 ± 26 MW m$^{-2}$ K$^{-1}$. By the use of C5OP linkers, the temperature across the interface became the 86% of the initial value when no linkers were adopted. Thus, we can assume that the reduction in thermal jump is responsible for the overall increase in thermal conductance, quantified in about 20% for C5OP. The temperature trend in biphenyl junction (Figure 7c) show a temperature jump reduced by approx. one third compared to the linker-less model, decreasing from 8.65 ± 0.57 K to 5.78 ± 0.40 K (Table 2). The temperature plot indicated a strong coupling between the graphene flakes, where the calculated temperature of the junction progressively matches the linear fit of the temperature slabs in graphene. Overall, for the three replicas, biphenyl jointed models reported a calculated thermal conductance of about 484 ± 55 MW m$^{-2}$ K$^{-1}$, 68% more than the linker-less model, mainly from the contribution of thermal jump reduction (67%, Table 2) than increased heat transferred, slightly over 12% (Table 1). Acene junction was also investigated with the purpose to find the upper-bound condition. The temperature of the junction as a function of the slabs displacement is reported in Figure 7d. In this case, a strongly reduced thermal jump (2.03 ± 0.32 K) is reflected into a superior thermal conductance (1363 ± 18 MW m$^{-2}$ K$^{-1}$), corresponding to about 4.7 times the thermal conductance of the linker-less junction.

*Table 2. Thermal jump, seed error and thermal jump reduction in polymer surrounded junctions for all the investigated linkers.*

| Model | Thermal jump [K] | | Ratio [%] |
| --- | --- | --- | --- |
| | Average value | Seed error $\frac{max - min}{2}$ | $\frac{linker\ type}{no\ linkers}$ |
| No linkers | 8.65 | ± 0.57 | 100 |
| C5OP | 7.46 | ± 0.31 | 86 |
| BP | 5.78 | ± 0.40 | 67 |
| ACN | 2.03 | ± 0.32 | 23 |

While the thermal conductance values calculated for the different PDMS embedded molecular junctions suggest large differences in their effectiveness, the conductance trend is in line with conductance values reported in self-standing junctions [12]. However, splitting the contribute in thermal transport from the PDMS chains and the molecular junctions represents a cahllenging issue, because the calculated thermal conductances are strongly dependent on the layout, in particular for the different heat transfer area, a parameter hard to quantify in molecules. By the analysis of thermostat heat flux from the suspended model of C5OP, BP and ACN, (with six molecules grafted, thus halved for this comparison), is about 0.05, 0.17 and 0.35 eV/ps, respectively. Table 3 reports the contribution from matrix (No Linkers) and the suspended molecules evaluated in our previous work [12] and embedded in PDMS in this chapter. From this analysis, the thermal contribution and matrix appears to be roughly cumulative, with value very close to the direct method calculation, made of polymer and junction, and the case where these constituents were taken separately.

*Table 3. Heat flux from direct method calculations (already reported in Table 1) and separate contribution from linkers and matrix*

| Model | Energy flux [eV/ps] | | | |
| --- | --- | --- | --- | --- |
| direct method | Direct method (average) | Thermostat error $\frac{max - min}{2}$ | Three suspended linkers | No Linkers + Three suspended linkers |
| No linkers | 1.47 | ± 0.07 | - | - |
| C5OP | 1.53 | ± 0.12 | + 0.05 ± 0.01 | 1.52 (-0.6%) |
| BP | 1.65 | ± 0.14 | + 0.17 ± 0.01 | 1.64 (-0.6%) |
| ACN | 1.78 | ± 0.11 | + 0.35 ± 0.02 | 1.82 (+2.2%) |

As an additional analysis, based on the very limited thermal jump across the acene-bound interface, the ACN junction may be considered equivalent to a continuous material, constituted by the two joined graphene sheets, thus neglecting the junction discontinuity. With this hypothesis, the temperature plot and linear fit depicted in Figure 8 was obtained.

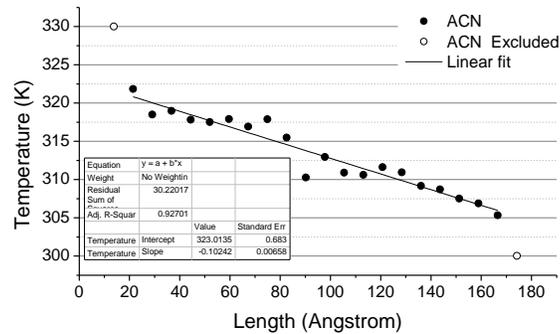

*Figure 8. Temperature of thermal slabs as a function of the position. The linear fit among Anthracene (ACN) slabs temperature suggests the suppression of the thermal jump across the junction.*

From the application of Fourier's law, the thermal conductivity of the joint graphene slabs was calculated at about 332 W m$^{-1}$ K$^{-1}$. This value may be considered representative of a network of graphene flakes fully joined by acene junctions within a PDMS matrix and represents a theoretical upper value for molecularly joined graphene nanocomposites.

## Conclusions

The thermal conductance between graphene platelets embedded in PDMS polymer matrix has been evaluated trough NEMD simulations. As expected from the study of the suspended junctions, the linker type influenced the overall conductance value also within the PDMS matrix. The aliphatic/aromatic C5OP junction yields a thermal conductance improvement of about 20% compared to the unbound sheets, while aromatic structures provided better enhancements, confirming trends previously reported for molecular junctions between self-standing graphene sheets. In particular, Biphenyl junctions enhanced thermal conductance by about 68% while anthracene linkers, considered as a sort of upper bound case study, yielded an almost 5-fold increase in thermal conductance. The strong dependence of the thermal conductivity in PDMS/graphene model is related to both the length and the chemistry of the linker. On one hand, short linkers forces the flakes to keep a smaller distance between the platelets than longer ones and consequently the volume of interposed polymer is smaller. On the other hand, the length of the linker is not enough to justify the observations and it is therefore suggested that the chemistry of the linker strongly affects the efficiency of the junction, even within the polymer matrix. Furthermore, the thermal transport across the contact between two conductive particles was found to correspond to the sum of heat transfer through the molecular junctions and through the PDMS molecules surrounding the particle/particle contact.

## Acknowledgements

This work has received funding from the European Research Council (ERC) under the European Union's Horizon 2020 research and innovation programme, Grant Agreement 639495 — INTHERM — ERC-2014-STG.

# References

[1] F. Müller-Plathe, A simple nonequilibrium molecular dynamics method for calculating the thermal conductivity, The Journal of chemical physics 106(14) (1997) 6082-6085.
[2] A. Khan, I. Navid, M. Noshin, H. Uddin, F. Hossain, S. Subrina, Equilibrium Molecular Dynamics (MD) Simulation Study of Thermal Conductivity of Graphene Nanoribbon: A Comparative Study on MD Potentials, Electronics 4(4) (2015) 1109-1124.
[3] A. Cao, Molecular dynamics simulation study on heat transport in monolayer graphene sheet with various geometries, Journal of Applied Physics 111(8) (2012) 083528.
[4] B. Mortazavi, S. Ahzi, Thermal conductivity and tensile response of defective graphene: A molecular dynamics study, Carbon 63 (2013) 460-470.
[5] B. Mortazavi, T. Rabczuk, Multiscale modeling of heat conduction in graphene laminates, Carbon 85 (2015) 1-7.
[6] T.M. Madkour, J. Mark, Polymer data handbook, Oxford University Press New York, 1999.
[7] C.S. Boland, U. Khan, G. Ryan, S. Barwich, R. Charifou, A. Harvey, C. Backes, Z. Li, M.S. Ferreira, M.E. Mobius, R.J. Young, J.N. Coleman, Sensitive electromechanical sensors using viscoelastic graphene-polymer nanocomposites, Science 354(6317) (2016) 1257-1260.
[8] S. Deng, V. Berry, Wrinkled, rippled and crumpled graphene: an overview of formation mechanism, electronic properties, and applications, Materials Today 19(4) (2016) 197-212.
[9] F. Liu, S. Song, D. Xue, H. Zhang, Folded structured graphene paper for high performance electrode materials, Adv Mater 24(8) (2012) 1089-94.
[10] J. Zhang, J. Xiao, X. Meng, C. Monroe, Y. Huang, J.M. Zuo, Free folding of suspended graphene sheets by random mechanical stimulation, Phys Rev Lett 104(16) (2010) 166805.
[11] H. Wang, H. Tian, S. Wang, W. Zheng, Y. Liu, Simple and eco-friendly solvothermal synthesis of luminescent reduced graphene oxide small sheets, Materials Letters 78 (2012) 170-173.
[12] A. Di Pierro, M.M. Bernal, D. Martinez, B. Mortazavi, G. Saracco, A. Fina, Aromatic molecular junctions between graphene sheets: a molecular dynamics screening for enhanced thermal conductance, RSC Advances 9(27) (2019) 15573-15581.